\documentclass[12pt]{article}

\usepackage[T2A]{fontenc}
\usepackage[cp1251]{inputenc}
\usepackage[english]{babel}

\usepackage[intlimits]{amsmath}
\allowdisplaybreaks
\usepackage{amsthm}
\usepackage{amssymb}

\newcommand\np{\mbox{NP}}
\newcommand\p{\mbox{P}}
\newcommand\uf{\mbox{UF}}
\newtheorem{theorem}{Theorem}
\newtheorem{corollary}[theorem]{Corollary}
\newtheorem{lemma}[theorem]{Lemma}

\author{Anatoly D. Plotnikov}
\title{On the structure of the class \np}
\date{The Dalh East-Ukrainian national university\\ Ukraine}

\begin{document}
\maketitle

\begin{abstract}
A new class \uf\ of problems is introduced, strictly included in the class \np, which arises in the analysis of the time verifying the intermediate results of computations. The implications of the introduction of this class are considered. First of all, we prove that $\p\not= \np$ and establish that it needs 
to consider the problem ``\p\ vs \uf'' instead the problem ``\p\ vs \np''. Also, we determine the set-theoretical of properties of a one-way functions that used in cryptology.
\end{abstract}

\section{Introduction}

A problem is defined as a certain general question that must be answered \cite{9}. {\it A mass problem} $Z$ is defined by means of the following information:
\begin{itemize}
\item by a parameter list;
\item by the properties of the problem solution.
\end{itemize}

A mass problem is also called simply {\it problem}. If the parameters of the problem have concrete values, then we have {\it instance}.

Several major classes of problems \cite{9, 41} are defined in complexity theory. Let us consider some of them.

A problem $Z$ belongs to the class \np\ if:
\begin{enumerate}
\item the problem can be defined by a finite number $n$ of symbols;
\item the problem solution can be represented by a finite number $m$ of symbols, 
where $m$ is a polynomial function of $n$: $m = f(n)$;
\item the time $t$ for verifying the obtained solution is some
polynomial functions of $n$: $t = \varphi(n)$.
\end{enumerate}

A set of problems of \np, having polynomial-time solution algorithm, forms a class \p, where $\p\subset \np$. An elementary example of a problem of the class \p\ is the problem of finding the maximum element of an array of size $n$. Clearly, for this it is necessary to consider all the elements of the array, i.e. the search time for a solution in this case equal to $O(n)$.

The problem $Z_1$ is {\it polynomially reduced} to the problem of $Z_2$ if each instance $X\in Z_1$ can be transformed in polynomial time in the instance $Y\in Z_2$ and a solution of $Y$ can be converted in polynomial time to a solution of $X$ \cite {41}.

In the class \np, universal or \np-complete problems are  allocated. The problem $Z$ is called {\it \np-complete} if any problem of \np\ reduced in polynomial time to the problem of $Z$. The set of all \np-complete problems forms the class NPC. There are hundreds of such problems in different fields \cite{9, compendium}.

Difficulty finding efficient (polynomial-time) algorithms for the \np-com\-ple\-te problems have led to the formulation of the problem ``\p\ vs \np'', whose solution is the answer to the question: how can we effectively solve \np-complete problems. In \cite {ch24} has been shown that there is a class of \np\ redundant.

There is also the problem of the existence of one-way function used in cryptology. This problem is also related to the solution of the previous problems ``\p\ vs \np''.

The aim of this paper to show the meaning and implications of the introduction of the new class of problems.

\section{Problems without foresight}

A Turing machine is a generally accepted model of the computing process for any problem of \np\ \cite {ch24}. The essence of this process is that the source data (parameters) of instance sequentially is converted per clock  cycle (step by step). The result of these changes is to obtain a solution of the problem, if it exists, or the answer that the solution of this problem does not exist. All the results of the computational process by Turing machine to obtain the final answer, obviously, should be considered as {\it intermediate results}.

In order to computational process was directed to obtain the correct result, it is necessary to check each step of computing an intermediate result with its belonging to some feasible solution of the problem.

For example, if we look for a solution of the satisfiability problem then at each step of choosing different literals (values of Boolean variables) to obtain a feasible solution, the results must satisfy the condition of consistency, i.e. impossible to choose the variable $x$ and its negation $\bar x$ in an intermediate solution simultaneously. And the possibility of such a choice in this problem is verified easily. 

In the definition of a problem of the class \np, verification procedure of an intermediate result does not defined. Therefore, in the class \np, we select a class of problems for which verification procedure of an intermediate result is the polynomial function of the problem dimension $n$. The set of all such problems is denoted \uf. Any problem of class \uf\ is called {\it a problem without foresight}.

The proposed name connected with the study of graph problem in the papers \cite{pl2, k11}, in which the result of the current selection of the solution depends on the subsequent election of such elements.

Thus, from the definition of the problem without foresight follows that in the problems that belong to $\np\setminus \uf$, the verification procedure of at least one intermediate result is an exponential function of the dimension of the problem. Here is another example of the optimization problem in the class $\np\setminus \uf$ as a decision problem.

{\bf Heavy tuple (HT)}

{\it Condition}. We are given a finite set $X=\{x_1, x_2, \ldots, x_n\}$ of Boolean variables, a bent-function $f(x_1, x_2, \ldots, x_n)\in \{0, 1\}$, the function $w(x_1, x_2, \ldots, x_n)$ $\in Z^{+}$ and the boundary $E\in Z^{+}$. The functions $f(X)$ and $w(X)$ are computed in the polynomial time.  

{\it Question}. Is there a set of values of Boolean variables $X$ such that $f(X)=1$ and the value of $w(X)$ is at least $E$?

\begin{lemma}
The HT problem belongs to the class \np.
\end{lemma}

{\bf Proof}. Indeed, the initial data of the HT problem and the solution are finite, and the solution has linear dimension of the problem dimension. In addition, the obtained tuple (solution) may be verified in the polynomial time. Therefore $HT\in \np$.\hfill Q.E.D.

\begin{lemma}
\label{tmain}
The HT problem can not be solved in polynomial time.
\end{lemma}

{\bf Proof}. In the HT problem, there is an implicit requirement to consider all tuples on which the bent-function $f(X)$ equals to 1. It follows of a fact that finding a tuple of maximum weight can only if we consider all the elements of an array of such tuples. Since one of the properties of a bent-function is that exactly on a half of the tuples $X$ its value equal to 1, and on the other half of tuples equal to 0, the verification of the intermediate results is possible only after considering the weights of all tuples on which $f(X)=1$. Since the number of such tuples is equal to $2^{n-1}$, from here it follows the validity of the above statement. \hfill Q.E.D.

\begin{theorem}
The HT problem belongs to the class $\np\setminus \uf$.
\end{theorem}

{\bf Proof}. It is easy to see that in the process of solving HT problem, intermediate result (a tuple $X$, where $f(X)=1$, and there is some value $w(X)$) requires at least $O(2^{n-1})$ time units to verify that the functions $w(X)$ satisfies the problem condition.\hfill Q.E.D.

\begin{corollary}
$\np\setminus \uf\not= \oslash$.
\end{corollary}

\begin{corollary}
\label{cor1}
$\uf\not= \np$.
\end{corollary}

\section{On the problem ``\p\ vs \np''}

The class \uf\ allows us to consider the problem ``\p\ vs \np'' and point to another proof mismatch of these classes than in cite{ch24}.

\begin{theorem}
If the problem $Z\in \np$ has a polynomial-time solution algorithm then $Z\in \uf$.
\end{theorem}

{\bf Proof}. Suppose that $Z\in \np\setminus \uf$. It follows that the intermediate results obtained in an exponential time, and the final result is obtained in a polynomial time. Contradiction, since the final result we obtain after the construction of intermediate results.  \hfill Q.E.D.

\begin{corollary}
\label{cor2}
$\p\subset \uf$.
\end{corollary}

\begin{theorem}
$\p\not= \np$.
\end{theorem}

{\bf Proof}. The validity of the above statement follows from Corollaries \ref{cor1} and \ref{cor2}.\hfill Q.E.D.

\section{On the existence of one-way functions}

The concept of an one-way function is used in cryptology. Suppose that a function $y=f(x)$ is one-way if the value $y$ for the given $x$ is computed ``easily'', and the value $x$ for the given $y$ is computed ''difficultly''. We will specify the concepts ''lightness'' and ''difficulty'' for an one-way function.

In computational complexity theory, the concept of ``easy'' or ``effective'' of computations means that the computation time $t$ of a function $y=f(x)$ is a polynomial function of the problem dimension $n$: $t=\lambda(n)$. Then the concept of ``difficult'' of computations means that the computation time $T$ of the function $x=f^{-1}(y)$, when it exists, is an exponential function of the problem dimension $n$: $T=\Lambda(n)$. Thus, we have the following result.

\begin{theorem}
If $y=f(x)$ is a one-way function then it belongs to the class $\p$, and the function $x=f^{-1}(y)$ exists and belongs $\np\setminus \uf$. 
\end{theorem}

{\bf Proof}. The validity of the above statement follows from the definition of a one-way function. \hfill Q,E,D.

\medskip
From the set-theoretic point of view, we can examine the function $y=f(x)$ as the mapping $R$: $Dom(R)\to Ran(R)$, where $x\in Dom(R)$ and $y\in Ran(R)$. Clear, that for every element $x\in Dom(R)$ of the mapping $x\to Ran(R)$ must contain the polynomial number of elements. On the other hand, the function $x=f^{- 1}(y)$ be the mapping $R^{- 1}$: $Dom(R^{- 1})\to Ran(R^{-1})$ or, equivalently, $Ran(R)\to Dom(R)$. In order that a function $y=f(x)$ was one-way, it is necessary, that the mapping of element $y\to Dom(R)$ has the exponential number of elements.
\medskip

An example of a one-way function is the bent-function $f(x_1,\ldots, x_n)$ of the HT problem. In this problem, the computation of ``direct'' function $f(X)$ and of ``weight'' $w(X)$ requires, by the condition, a polynomial time of $n$, and the inverse problem --- search tuple of the maximum weight when $f(X)=1$ --- requires an exponential time of $n$

Thus, we have proved the following statement.

\begin{theorem} 
\label{rmain}
A class $Q\subset \np$ of one-way functions is not an empty, and each one-way function $y=f(x)$ has the following properties:
\begin{itemize}
\item an argument $x$ can be expressed by a finite number of symbols $n$;
\item a value of $y$ can be written $m$ symbols, where $m$ is of a polynomial function of $n$: $m=\pi(n)$;
\item the computation time $t$ of the function $y$ is a polynomial function of $n$: $t=\varphi(n)$;
\item the computation time of some intermediate values of $x=f^{-1}(y)$ (or its length) is an exponential function of $n$.
\end{itemize}
\end{theorem}

\section{Conclusion}
 
In our view, the formal solution ``\p\ vs \np'' found in the narrow sense. But only one step has made to solving the problem of solvability for a class \np\ in the broad sense. This result shows that the accepted definition of the class \np\ is redundant and can be removed from the consideration of the problems of $\np\setminus \uf$. In other words, consider the polynomial solvability of the class \uf, i.e. instead the problem ``\p\ vs \np'' we have a problem ``\p\ vs \uf''. This problem can be solved if we find a polynomial-time algorithm for solving a \np-complete problem $Z\in NPC$ or it is proven that a polynomial-time algorithm for solving \np-complete problem does not exist.

In \cite{algebra1} proposed the heuristic polynomial-time algorithm for solving \np-complete problem for the maximum independent set problem. This algorithm is based on the construction of a directed graph that satisfies certain conditions that allows to find a maximal independent set as a maximum antichain poset. There have not yet found a counterexample for the proposed algorithm.. 

From the theory of \np-completeness, it follows that any problem of the class \np\ can be reduced to any \np-complete problem in a polynomial time. This raises a number of questions, since it is known that any reduction allows us to restate the difficulties of solving one problem in difficulty of other problem.

Also, it follows from the paper that a cryptosystem is theoretically stable if it satisfies Theorem \ref {rmain}, when it not based on the difficulties of working with huge numbers.

\end{document}